\documentclass[12pt]{article}
\usepackage{times,latexsym}

\def\half{{\textstyle \frac{1}{2}}}%

\newcommand{\beq}{\begin{equation}}%
\newcommand{\eeq}{\end{equation}}%
\newcommand{\beqs}{\begin{eqnarray}}%
\newcommand{\eeqs}{\end{eqnarray}}%

\def\pdr{{\partial}}

\begin{document}
\title{\bf  Dissipative Mechanics Using Complex-Valued Hamiltonians}
\author{ S. G. Rajeev \\
Department of Physics and Astronomy\\
        University of Rochester, Rochester, New York 14627\\
	}
\maketitle
\begin{abstract}
We show that a large class of dissipative systems can be brought to a canonical form by introducing complex co-ordinates in phase space and a complex-valued hamiltonian. A naive canonical quantization of these systems lead to non-hermitean hamiltonian operators. The excited states are unstable and decay to the ground state . We also compute the tunneling amplitude across a potential  barrier. 

\end{abstract}



\newpage
\section{\bf Introduction}

In many physical situations, loss of energy of the system under study
to the outside environment cannot  be ignored. Often, the long time
behavior of the system is determined by this loss of energy, leading
to interesting phenomena such as attractors.

There is an extensive literature on  dissipative systems at both the
classical and quantum levels (See for example the  textbooks
\cite{UlrichWeiss,WolfgangSchleich,ScullyZubairy}). Often the theory is based on an evolution equation of the density matrix of a `small system' coupled
to a `reservoir' with a large number of degrees of freedom, after the
reservoir has been averaged out. In such approaches  the
system is described  by a mixed state rather than a pure state: in quantum mechanics
by a density instead of a wavefunction and in classical mechanics by a
density function  rather than a point in the phase space.

There are other approaches that do deal with the evolution equations of a pure  state.
The  canonical formulation of classical mechanics does not apply in a
direct way to dissipative systems because the hamiltonian usually has the
meaning of energy and would be  conserved. By redefining the Poisson
brackets \cite{newPB} , or by using time dependent hamiltonians
\cite{timedepham}, it is possible to bring such  systems within a canonical
framework. Also, there are   generalizations of the Poisson bracket that may not
be anti-symmetric and/or may not  satisfy the Jacobi identity
\cite{PMorrison,Marsden} which give dissipative equations.

We will follow another route, which turns out in many cases to be  simpler than the above. It is suggested by  the simplest example, that of
the damped simple harmonic oscillator.
 As is well known, the effect of damping is to replace the natural
 frequency of oscillation by a complex number, the imaginary part of
 which determines   the rate of exponential decay of energy. Any
 initial state  will decay to the ground state (of zero energy) as time tends to infinity.
The corresponding co-ordinates in phase space (normal modes) are
complex as well. This suggests that the equations are of hamiltonian
form, but with a complex-valued hamiltonian.

It is not difficult to verify that this is true directly. The real part of the
hamiltonian is a harmonic oscillator, 
although with a shifted frequency; the imaginary part is its constant
multiple.  If we pass to the quantum theory in the usual way, we get a
non-hermitean hamiltonian operator. 
Its eigenvalues are complex valued, except for the ground state which
can be
 chosen to have a real eigenvalue. Thus all states except the ground
 state are unstable.
 Any state decays to its projection to the ground state as time tends
 to infinity. This is a reasonable quantum analogue of the classical decay of energy.

We will show  that a wide class of dissipative systems
can be brought to such a canonical form using  a complex-valued
hamiltonian.
The usual  equations of motion determined  by a hamiltonian and Poisson bracket are
\beq
{d\over dt}\pmatrix{p\cr x}=\pmatrix{\{H,p\}\cr \{H,x\}}.
\eeq
 At first a complex-valued function 
\beq
{\cal H}=H_1+iH_2
\eeq
does not seem to make sense when put into the above formula:
\beq
{d\over dt}\pmatrix{p\cr x}=\pmatrix{\{H_1,p\}\cr \{H_1,x\}}+i
\pmatrix{\{H_2,p\}\cr \{H_2,x\}}
\eeq
since the l.h.s. has real components.
How can we make sense of multiplication by $i$ and still get  a vector with only real components?

Let us consider a complex number $z=x+iy$ as an ordered  pair of real 
numbers $(x,y)$. The effect of multiplying $z$ by $i$ is the linear transformation 
\beq
\pmatrix{x\cr y}\mapsto \pmatrix{-y\cr x}
\eeq
on its components. That is, multiplication by $i$ is equivalent to the action by the matrix 
\beq
J=\pmatrix{0&-1\cr 1&0}.
\eeq
Note that $J^2=-1$. Geometrically, this corresponds to a rotation by ninety degrees.

Generalizing this, we can interpret multiplication by $i$ of a vector field 
in phase space to mean the action by some matrix  $J$
satisfying 
\beq
J^2=-1. \label{almostcomplex}
\eeq
Given such a matrix, we can define the equations of motion  generated by a complex-valued function ${\cal H}=H_1+iH_2$ to be
\beq
{d\over dt}\pmatrix{p\cr x}=\pmatrix{\{H_1,p\}\cr \{H_1,x\}}+J
\pmatrix{\{H_2,p\}\cr \{H_2,x\}}
\eeq
Our point  is that the infinitesimal time evolution of a  wide class of mechanical systems is of this type for an appropriate choice of $\{,\}, J,H_1$ and $H_2$. 

In most cases there is a complex-cordinate system in which $J$ reduces to a simple multiplication by $i$; for example on the plane this is just $z=x+iy$. For such a co-ordinate system to exist the tensor field has to satisfy certain integrability conditions in addition to (\ref{almostcomplex}) above. These conditions are automatically satisfied if the matrix elements of $J$  are constants.

What would be the advantage of fitting dissipative systems into such a 
complex canonical formalism? A practical advantage is that they can lead to better numerical approximations, generalizing the  symplectic integrators widely used  in hamiltonian systems: these integrators preserve the geometric structure of the underlying physical  system. Another is that it allows us to use ideas from hamiltonian mechanics to study structures unique to dissipative systems such as strange attractors.  We will not pursue these ideas in this paper.

Instead we will look into the canonical quantization of dissipative systems. The usual correspondence principle leads to a non-hermitean hamiltonian. As in the elementary example of the damped simple harmonic oscillator, the eigenvalues are complex-valued. The excited states  are unstable and decay to the ground state. Non-hermitean hamiltonians have arised already in several dissipative systems in condensed matter physics \cite{DavidNelson} and in particle physics \cite{KKbar}. The Wigner-Weisskopf approximation provides a physical justification for using a non-hermitean hamiltonian.
A dissipative system is modelled by coupling it to some other `external' degrees of freedom so that the total hamiltonian is hermitean and is conserved.
In second order perturbation theory we can eliminate the external degrees of freedom to get an effective hamiltonian that is non-hermitean. 

It is interesting  to compare our approach with the tradition of  Caldeira and Leggett\cite{CaldeiraLeggett}. Dissipation is modelled by coupling the original (`small') system to a thermal bath of harmonic oscillators. After integrating out the oscillators in the path integral formalism an effective action for the small system is obtained. A complication is that this effective action is non-local: its extremum ( which dominates tunneling) 
is the solution of an integro-differential equation. We will see that the 
integral operator appearing here is also a complex structure (the Hilbert transform), although one non-local in time and hence different from our use of complex structures.

We calculate the tunneling amplitude of a simple one dimensional quantum system within our framework. Dissipation can increase the tunneling probability, which is not allowed in the Caldeira-Leggett model. 

We begin with a brief review of the most elementary case, the damped simple harmonic oscillator. Then we generalize to the case of a generic one dimensional system with a dissipative  force proportional to velocity. Further generalization  to systems with several degrees of freedom is shown to be possible provided that the dissipative force is of the form
\beq
-\pdr_a\pdr_bW {dx^b\over dt}
\eeq
for some function $W$. In simple cases this function is just the square of the distance from the stable equilibrium point. Finally, we show how to bring a dissipative system whose configuration space is a Riemannian manifold into this framework. This is important to include interesting systems such as the rigid body or a particle moving on  a curved surface. We hope to return to these examples in a later paper.

\section{ Dissipative Simple Harmonic Oscillator}
We start by recalling the most elementary example of a classical dissipative system, described  the differential equation
 \beq
 \ddot x+2\gamma\dot x +\omega^2x=0, \gamma>0.
 \eeq
 We will consider the under-damped case  $\gamma<\omega$ so that the system is still oscillatory.
 
We can write these equations in phase space 
\beqs
\dot x&=&p\\
\dot p&=&-2\gamma p-\omega^2 x
\eeqs
The energy
\beq
H=\half[p^2+\omega^2 x^2]
\eeq
decreases monotonically along the trajectory:
\beq
{dH\over dt}=p\dot p+\omega^2 x\dot x=-2\gamma p^2\leq 0.
\eeq
The only trajectory which conserves energy is the one with $p=0$, which must have $x=0$ as well to satisfy the equations of motion.

These   equations can be brought to diagonal form by a linear transformation:
\beq
z=A\left[-i(p+\gamma x)+\omega_1 x\right],\quad {dz\over dt}=[-\gamma+i\omega_1]z
\eeq
where
\beq
 \omega_1=\sqrt{\omega^2-\gamma^2}.
\eeq
The  constant $A$ that can be chosen later  for  convenience. These complex co-ordinates are the natural variables (normal modes) of the system.

\subsection{Complex Hamiltonian}
We can  think of the DSHO as a generalized hamiltonian system with a complex-valued hamiltonian.

The Poisson bracket $\{p,x\}=1$  becomes, in terms of the variable $z$,
\beq
\{z^*,z\}=2i\omega_1 |A|^2
\eeq
So if we choose $A={1\over \sqrt{2\omega_1}}$
 \beq
\{z^*,z\}=i
\eeq
 So the   {\bf complex-valued} function 
 \beq
{\cal H}=(\omega_1+i\gamma) zz^*.
\eeq
 satisfies
\beq
{dz\over dt}=\{{\cal H},z\},\quad {dz^*\over dt}=\{{\cal H}^*,z^*\}
\eeq

Of course, the limit $\gamma\to 0$ this ${\cal H}$  tends to the usual hamiltonian $H=\omega zz^*$.
Thus, on any analytic function $\psi$, we will have
\beq
{d\psi\over dt}=\{{\cal H},\psi\}=[\omega_1+i\gamma]z{\pdr \psi\over \pdr z}
\eeq

\subsection{Quantization}

By the usual rules of canonical quantization,  the  quantum theory is given by turning  ${\cal H}$ into a
non-hermitean operator by replacing $z\mapsto a^\dag,\ z^*\mapsto
\hbar a$ and 
\beq
[a,a^\dag]=1,\quad a^\dag=z,\quad  a ={\pdr \over \pdr z},\quad {\cal
  H}=\hbar (\omega_1+i\gamma) a^\dag a.
\eeq
The
effective hamiltonian ${\cal H}=H_1+iH_2$  is normal (
i.e., its hermitean and anti-hermitean parts commute, $[H_1,H_2]=0$ ) so it is still meaningful to speak of eigenvectors of ${\cal H}$. The eigenvalues are complex 
\beq
(\omega_1+i\gamma) n, n=0,1,2,\cdots.
\eeq
 The higher excited states are  more and more unstable. But the ground state is stable, as its eigenvalue is zero. 

Thus a generic state
\beq
\psi=\sum_{n=0}^\infty  \psi_n |n>
\eeq
 will evolve in time as 
\beq
\psi(t)=\sum_{n=0}^\infty  \psi_n e^{i\hbar [\omega+i\gamma]nt}|n>.
\eeq
Unless $\psi$ happens to be orthogonal to the ground state $|0>$,  the
wavefunction will tend  to the ground state as time tends to
infinity;  final state will be the projection of the initial state
to the ground state. This is the quantum analogue of the classical
fact that the system will decay to the minimum energy state as time
goes to infinity. All this sounds physically reasonable.

\subsection{ The Schr\"odinger Representation}

In the Schr\"odinger representation,this amounts to
\beq
a={1\over \sqrt{2\hbar \omega_1}}\left[\omega_1 x+\hbar {\pdr \over
    \pdr x}\right],
\quad
a^\dag={1\over \sqrt{2\hbar \omega_1}}\left[\omega_1 x-\hbar {\pdr \over \pdr x}
\right]
\eeq
\beq
\hat{\cal H}=\left(1+i{\gamma\over \omega_1}\right)\left[-{\hbar^2\over 2}{\pdr^2\over \pdr x^2}+\half\omega_1^2 x^2-\half\hbar\omega_1\right]
\eeq
Thus the operator representing momentum $p$ is
\beq
\hat p=-i\hbar{\pdr \over \pdr x}-\gamma x
\eeq
which includes a subtle correction dependent on the friction.

  The
time evolution operator can be chosen to be 
\beq
\hat{\cal H}_{\rm Schr}=\hat H+\hat H_{\rm diss}
\eeq
where 
\beq
\hat H=-{\hbar^2\over 2}{\pdr^2\over \pdr x^2}+\half\omega^2 x^2
\eeq
is the usual harmonic oscillator hamiltonian and
\beq
\hat H_{\rm diss}=-\half\gamma^2 x^2+i{\gamma\over \omega_1}\left[-{\hbar^2\over
    2}{\pdr^2\over \pdr x^2}+\half{\omega_1^2}
  x^2-\half\hbar\omega_1\right]
\eeq
This is slightly different from the operator  $\hat {\cal H}$ above, because the ground
state energy is not fixed to be zero. The constant in $H_{\rm
  diss}$ has been chosen so that this state has zero imaginary part
for 
its eigenvalue.

\section{ Dissipative System of One Degree of Freedom}

We will now generalize to a  non-linear  one-dimensional oscillator   with friction:
\beq
 {dp\over dt}=-{\pdr V\over \pdr x}-2\gamma p,\quad {dx\over dt}=p,\quad \gamma>0.
\eeq
The DSHO is the special case $V(x)=\half \omega^2x^2$.
The idea is that we lose energy whenever the system is moving, at a rate proportional to its velocity.  It again follows that 
\beq
{dH\over dt}=-2\gamma p^2\leq 0
\eeq
where $H=\half p^2+V$. These equations  can be written as 
\beq
{d\xi^i\over dt}=\{H,\xi^i\}-\gamma^{ij}\pdr_jH
\eeq
where $\gamma=\pmatrix{2\gamma&0\cr 0&0}$ is a positive but degenerate matrix.

\subsection{Complex Effective Hamiltonian}

In the case of the DSHO, we saw that it is the combination
\beq
p_1=p+\gamma x.
\eeq
rather than $p$ that appears naturally ( for example in the normal co-ordinate $z$). 
In terms of  $(p_1,x)$  the equations above  take the form
\beq
 {dp_1\over dt}=-{\pdr V_1\over \pdr x}-\gamma p_1,\quad {dx\over dt}=p_1-\gamma x,
\eeq
where
\beq
V_1(x)=V(x)-\half\gamma^2 x^2;
\eeq
i.e.,
\beq
{d\over dt}\pmatrix{p_1\cr x}=\pmatrix{\{H_1,p\}\cr \{H_1,x\}}-\gamma\pmatrix{p_1\cr x}.
\eeq

We would like to see if we can write these as canonical equations of motion with a complex-valued effective hamiltonian. Suppose we define
\beq
H_2={\gamma\over 2\omega_2}[p^2+\omega_2^2x^2],\quad J=\pmatrix{0&-\omega_2\cr {1\over \omega_2}&0}
\eeq
for some constant $\omega_2$ which we will choose later. 
Note that $J$ is a complex structure; i.e., 
\beq
J^2=-\pmatrix{1&0\cr 0&1}.
\eeq
Then the equations of motion take the form
\beq
{d\over dt}\pmatrix{p_1\cr x}=\pmatrix{\{H_1,p_1\}\cr \{H_1,x\}}+ J\pmatrix{\{H_2,p_1\}\cr \{H_2,x\}}.
\eeq
In terms of the complex co-ordinate
\beq
z={1\over \sqrt{2\omega_2}}[\omega_2 x-ip_1]
\eeq
this becomes
\beq
{dz\over dt}=\{H_1+iH_2,z\}.
\eeq

Thus the non-linear oscillator also can be written as a canonical system with a complex valued hamiltonian ${\cal H}=H_1+iH_2$; with 
 $\{H_1,H_2\}\neq 0$ in general. But there are many ways to do this, parametrized by $\omega_2$.  The natural choice is  
 \beq
 \omega_2=\sqrt{V''(x_0)-\gamma^2},
 \eeq
  where $x_0$ is the equilibrium point at which $V'(x_0)=0$. Then, in the neighborhood of the equilibrium point the complex structure  reduces to  that of the DSHO.

\subsection{Quantization of a Dissipative One-dimensional System}

We can quantize the above system by applying the usual rules of canonical quantization to the complex effective hamiltonian $H_1+iH_2$ to get the operator:
\beq
\hat{\cal H}=-{\hbar^2\over 2}{\pdr^2\over \pdr x^2}+V(x)-\half\gamma^2x^2+i{\gamma\over \omega_2}\left[
-{\hbar^2\over 2}{\pdr^2\over \pdr x^2}+\half\omega_2^2x^2+c
\right].
\eeq
The Schr\"odinger equation 
\beq
-i\hbar{\pdr \psi\over \pdr t}=\hat{\cal H}\psi
\eeq
then determines the time evolution of the dissipative system.

The anti-hermitean part of the hamiltonian is bounded below so the imaginary part of the eigenvalues will be bounded below.
We can choose the real constant $c$ such that the eigenvalue with the
smallest imaginary part is actually real. Then the generic state will
evolve to this ` ground state'.

\vfill\break
{\bf Remarks}\hfill\break
\begin{enumerate}

\item Our model of dissipation amounts to adding  a  term
\beq
\hat H_{\rm diss}=-\half\gamma^2x^2+i{\gamma\over \omega_2}\left[
-{\hbar^2\over 2}{\pdr^2\over \pdr x^2}+\half\omega_2^2x^2
\right]+c
\eeq
to the usual conservative hamiltonian
\beq
\hat H=-{\hbar^2\over 2}{\pdr^2\over \pdr x^2}+V(x).
\eeq
The dissipative term we add is very close to being  anti-hermitean;
i.e., except for  the term $-\half\gamma^2 x^2$ which is second order in the dissipation.

\item The Schr\"odinger equation for time reversed QM ( evolving to the past rather than future) is ${\cal H}^\dag$. The eigenvalues of ${\cal H}^\dag$ are the complex conjugates of those of ${\cal H}$, but the eigenfunctions may not be  the same as those of ${\cal H}$ since $[{\cal H},{\cal H}^\dag]\neq 0$ in general; i.e., the operator ${\cal H}$ may not be normal. ( For the DSHO the commutator vanished so the issue did not arise.) Moreover the eigenfunctions of ${\cal H}$ corresponding to distinct eigenvalues need not be orthogonal; they would  still  be linearly independent, of course. 

 There are examples of such non-normal hamiltonians in  nature when  time-reversal invariance is violated ( e.g., $K\bar K$ oscillation)\cite{KKbar}.  But in ordinary quantum mechanics such loss of time reversal invariance might be  unsettling. Carl Bender \cite{CarlBender} has suggested that for such non-hermitean hamiltonians, real eigenvalues and time reversal invariant dynamics can be recovered by modifying the inner product in the Hilbert space. This is the quantum counterpart to the reformulation of 
 classical DSHO as a conservative canonical system by modifying the Poisson bracket and hamiltonian \cite{newPB}.
 
 However we note that in the presence of dissipation, the classical equations of motion  are not time reversal invariant either: energy would grow rather than dissipate. So we should not expect quantum mechanics of dissipative systems to be time reversal invariant either.
 The appropriate symmetry is 
 \beq
 [{\cal H}(\gamma)]^\dag={\cal H}(-\gamma)
 \eeq
 which is satisfied in our case.

\item Note that within our model, the anti-hermitean part of the hamiltonian is  a sort of  harmonic oscillator even if the hermitean part has non-linear classical dynamics. This is because we chose a particularly simple form of dissipation, $\dot p\sim -\gamma p$. If we had chosen a more complicated ( e.g., non-linear) form of dissipation, the anti-hermitean part would be more complicated. It is common to model a dissipative system by coupling it to a thermal bath of oscillators. The dissipation is determined by the spectral density of the frequencies of these oscillators. Each choice of spectral density leads to a different dissipation term and, in our description, to a different anti-hermitean part for the hamiltonian. But if the dissipative force  is small it is reasonable to expect that it is linear in the velocity.

\end{enumerate}

\section{ Comparison with the Approach of Caldeira and Leggett}

It is interesting to see how our approach compares with the
influential work of Caldeira and Leggett \cite{CaldeiraLeggett}

Their basic idea is to couple the one dimensional system to a
collection of harmonic oscillators; the transfer of energy to the
oscillators leads to its dissipation in  the original  system.
The lagrangian is\footnote{ We change notation slightly from them to
  avoid conflict with ours.}
\beq
L=\half \dot x^2-V(x)+\sum_\alpha[\half m_\alpha \dot q_\alpha^2-\half
m_\alpha \omega_\alpha^2 q_\alpha^2] -x\sum_\alpha C_\alpha q_\alpha
-\half \Delta\omega^2 x^2.
\eeq
In the last term, $\Delta\omega^2$ is chosen such that the equilibrium point of
the variable $x$ is still the minimum of $V$: it is a finite
renormalization of the potential induced by the coupling to the oscillators.

 In the euclidean  path
integral approach, they derive an effective action after integrating
out these oscillators:
\beq
S_{\rm eff}[x]= \int_0^T[\half \dot x^2+V(x)]d\tau+\half\int_{-\infty}^\infty d\tau'\int_0^T d\tau [x(\tau)-x(\tau')]^2 a(\tau-\tau')
\eeq
where
\beq
a(\tau)=\sum_\alpha {C_\alpha^2\over 4m_\alpha
  \omega_\alpha}e^{-\omega_\alpha\tau}\equiv {1\over 2\pi}\int_0^{\infty} \rho(\omega)e^{-\omega\tau}d\omega
\eeq
All the properties of the oscillators that matter to us are contained
in the function $\rho(\omega)$. The choice $\rho(\omega)=\gamma\omega$
corresponds to  the usual dissipative term . This at first sounds
strange: how can this effective action that is non-local in time be
related to the differential equation of the damped oscillator?
The minimum of $S_{\rm eff} $  ( which determines the tunneling
amplitude  in the leading WKB approximation) is given by the
integro-differential equation\footnote{ Here, ${\cal P}$ stands for the
  Cauchy Principal part.}
\beq
{d^2 x\over d\tau^2}={\pdr V\over \pdr x}+{\gamma\over \pi}{\cal P}\int_{-\infty}^\infty
d\tau' {x(\tau)-x(\tau')\over (\tau-\tau')^2}.
\eeq
 In
the absence of damping we would get this equation for the most likely
path in the classically forbidden region  by the replacement
\cite{CallanColeman} :
\beq
t\mapsto i\tau
\eeq
in  Newton's law:
\beq
{d^2 x\over dt^2}=-{\pdr V\over \pdr x}.
\eeq
Such a naive replacement would not make sense in the damped equation:
\beq
{d^2 x\over dt^2}=-{\pdr V\over \pdr x}-\gamma{dx\over dt}\mapsto 
{d^2 x\over d\tau^2}={\pdr V\over \pdr x}-i\gamma{dx\over d\tau}.
\eeq
Since $x$ is real, the multiplication by $i$ on the r.h.s. does not
make sense.

Perhaps
multiplication by $i$ should be replaced by the action of some linear
operator on the space of functions of one variable? If so the square
of this linear operator must be $-1$.  There is a well-known operator
with this property, the Hilbert transform \cite{HilbertTransform}:
\beq
{\cal J}  x(t)={1\over \pi}{\cal P}\int_{-\infty}^\infty {x(t')\over t-t'} dt'
\eeq
The Caldeira-Legget equation of motion is obtained by replacing the
$i$ in the naive equation above by the operator  ${\cal J}$:
\beq
{d^2 x\over d\tau^2}={\pdr V\over \pdr x}-\gamma{d\over d\tau}{\cal J}x.
\eeq

Thus although obtained by very different arguments, there is a way to
understand the Caldeira-Legget equation of motion in terms of a
complex structure ( the Hilbert transform) on the space of paths. We instead introduce a
complex structure in the phase space at the classical level. Could
ours be the canonical quantization of their non-local path integral?
\section{ Tunneling in a Dissipative System}

Perhaps the most interesting question about  quantum dissipative systems
is how dissipation affects tunneling. The standard WKB approximation method adapts easily to our case. 

For illustrative purposes, it suffices to  consider a one-dimensional system with a potential
\beq
V(x)=\half \omega^2x^2\left(1-{x\over a}.
\right)
\eeq
We ask for the tunneling probability amplitude from the origin $x=0$ to the
point $x=a$ in a long time. In the absence of dissipation this is
given in the WKB approximation by 
\beq
e^{-{1\over \hbar}\int_0^a \sqrt{2V(x)} dx}.
\eeq 
The integral in the exponent is the minimum of the imaginary time action
\beq
\int_{0}^\infty[\half \dot x^2+V(x(\tau))]d\tau
\eeq
among all paths satisfying the boundary conditions $x(0)=0$ and
$x(\infty)=a$. (This minimizing path is the `instanton'.)

If we apply the WKB approximation to the Schr\"dinger equation we get 
\beq
\psi=e^{-{\phi\over \hbar}},\quad -\half\left({\pdr\phi\over \pdr x}\right)^2+V_1(x)
+i{\gamma\over \omega_2}\left[-\half\left({\pdr\phi\over \pdr
      x}\right)^2+\half\omega_2^2 x
\right]\approx 0.
\eeq
Solving this,  
\beq
\phi(x)=\int_0^x\sqrt{2V_1(x)+i\gamma\omega_2 x^2\over 1+i{\gamma\over \omega_2}
}dx
\eeq
The tunneling probability  is given by 
\beq
e^{-2{\rm Re\ }\phi(b)}
\eeq
where $b$ is the point of escape from the potential barrier; it might
depend on the dissipation.


It looks most natural to choose
$\omega_2=\omega_1=\sqrt{\omega^2-\gamma^2}$ as in the case of the
DSHO. Then, 
\beqs
\phi(x)&=&\omega_1\int_0^x \sqrt{1-{\omega^2\over \omega_1(\omega_1+i\gamma)}{x\over a}}\  \ xdx.
\eeqs
Now,
\beq
\int_0^b x\sqrt{b-x}dx={4\over 15}b^{5\over 2}.
\eeq
There are small discrepancies ( up to higher order terms in $\gamma$) depending on whether we think  of the
point of escape as $a$, or ${\omega_1^2\over \omega^2}a$ or the
complex number ${\omega_1(\omega_1+i\gamma)\over \omega^2}a$ which is the zero of the integrand. The last
choice gives the simplest answer for the tunneling probability
\beq
e^{-{8\over 15}{a^2\omega^2\over \hbar \omega}{(\omega^2-\gamma^2)^{3\over
    2}(\omega^2-2\gamma^2)\over \omega^5} }
\eeq
 This differs from  the results of Caldeira and
Legget: the tunneling probability is enhanced by 
dissipation. There could still  be systems in nature that are described by our model.

\section{Multi-Dimensional Dissipative Systems}

Now we will generalize to a multi-dimensional dynamical system while also allowing for a certain kind of non-linearity in the frictional force.

Suppose  the hamiltonian is the sum of a kinetic and a potential energy,
\beq
H=\half p_ap_a+V(x).
\eeq
With the usual Poisson brackets
\beq
\{p_a,x^b\}=\delta_a^b,\quad \{p_a,p_b\}=0=\{x^a,x^b\}
\eeq
the conservative equations of motion are
\beq
{dp_a\over dt}=-{\pdr_a} V,\quad {dx^a\over dt}=p^a
\eeq

We now assume that the frictional force is of the form $-2\gamma_{ab}\dot x^b$ for some  positive  non-degenerate tensor $\gamma_{ab}$ that might depend on  $x$.
The idea is that the system loses energy when parts of it move relative to other parts or relative to some medium in which the system is immersed. Then the equations of motion become
\beq
{dp_a\over dt}=-{\pdr_a} V-2\gamma_{ab} p_b,\quad {dx^a\over dt}=p^a.
\eeq
Here  we raise and lower indices using the flat euclidean  metric $\delta_{ab}$.
So
\beq
{dH\over dt}=-2\gamma^{ab}p_ap_b\leq 0.
\eeq
If the dissipation tensor  happens to be the  Hessian of some  convex function
\beq
\gamma_{ab}=\pdr_a\pdr_bW,
\eeq
it is possible to write these as Hamilton's equations with a complex-valued hamiltonian.

Then, we would have
\beq
{d\over dt}\pdr_aW=\gamma_{ab}p^b,\quad \gamma_{ab}\pdr_bW=\pdr_a(\half\pdr_bW\pdr_bW).
\eeq
Using these identities we can rewrite the equations of motion in the new variables:
\beq
\tilde p_a=p_a+\pdr_aW
\eeq
as
\beq
{d\tilde p_a\over dt}=-{\pdr_a} [V-\half (\pdr W)^2]-\gamma_{ab}\tilde p_b,\quad {dx_a\over dt}=\tilde p_a-\pdr_aW.
\eeq

This motivates us to define
\beq
H_1=\half \tilde p^2+V-\half(\pdr W)^2,\quad H_2={1\over \omega_2}[\half\gamma_{ab}\tilde p_a\tilde p_b+\omega_2^2 W],\quad J=\pmatrix{0&-\omega_2\cr {1\over \omega_2}&0}
\eeq
for some positive constant $\omega_2$.
Note that $J$ is a complex structure, $J^2=-1$.
Then
\beq
{d\over dt}\pmatrix{\tilde p\cr x}=\pmatrix{\{H_1,\tilde p\}\cr \{H_1,x\}}+J\pmatrix{\{H_2,\tilde p\}\cr \{H_2,x\}}
\eeq
In terms of the complex variable 
\beq
z_a={1\over \sqrt{ 2\omega_2}}[\omega_2 x_a-i\tilde p_a]
\eeq
this is just 
\beq
{dz_a\over dt}=\{H_1+iH_2,z_a\}
\eeq
just as before. Thus once the dissipation tensor is of the form $\gamma_{ab}=\pdr_a\pdr_bW$ the whole framework generalizes easily. The effect of dissipation is to add the term 
\beq
H_{\rm diss}=-\half(\pdr W)^2+{i\over \omega_2}[\half \gamma_{ab}p_ap_b+\omega_2^2W]
\eeq
to the hamiltonian. The classical equations turn out to be independent of the choice of $\omega_2$, but the quantum theory will depend on its choice.

\subsection{Quantization}

We can then quantize this operator  in the Schr\"odinger picture:
\beq
\hat {\cal H}=\hat H+\hat H_{\rm diss},
\eeq
where
\beq
\hat H\psi=-\half\nabla^2\psi+V\psi,\quad H_{\rm diss}\psi=-\half (\nabla W)^2\psi+{i\over \omega_2}\left[-{1\over 2}\pdr_a\left(\gamma^{ab}\pdr_b\psi
\right)
+\omega_2^2W
\right]
\eeq

A technical point to note here is that there are two competing metrics in the story. The kinetic energy is $\half p_ap_a$, determined by the euclidean metric. But the quadratic term in the dissipative part of the hamiltonian is determined by some other tensor $\gamma_{ab}$. We have chosen to use the Euclidean metric $\delta_{ab}$ to define derivatives and to raise and lower indices. 
Thus, $\gamma^{ab}=\delta^{ac}\delta^{bd}\gamma_{cd}$  and not the inverse of $\gamma_{ab}$.

 The operator $\pdr_a\left(\gamma^{ab}\pdr_b\psi\right)$ is thus a kind of `mixed' laplacian that uses $\gamma_{ab}$ as well as, implictly, the euclidean metric. Since $\pdr_a\gamma^{bc}\neq 0$ in general, the ordering of factors is important. with the order we chose,$\pdr_a\left(\gamma^{ab}\pdr_b\psi\right)$ is hermitean and positive.

\section{Geometric Model of Dissipative  Mechanics}

We will now further generalize to the case of dissipative  dynamics on a cotangent bundle $T^*Q$.

The hamiltonian is again the  sum of  kinetic  and potential energies
\beq
H=\half g^{ab}p_ap_b+V(x).
\eeq
except that we now allow the tensor $g^{ab}$ not to be constant. We
will use $g_{ab}$ ( the inverse of $g^{ab}$)  as the Riemann metric,
 used to define covariant derivative $\nabla_a$  and to raise and lower indices.

The conservative equations of motion are
\beq
{dp^a\over dt}+\Gamma^a_{bc}p^bp^c=-g^{ab}{\pdr_b} V,\quad {dx^a\over dt}=p^a
\eeq
where $\Gamma^a_{bc}$ is the usual Christoffel symbol.

With friction added,
\beq
{dp^a\over dt}+\Gamma^a_{bc}p_b p^c=-g^{ab}{\pdr_b} V-2\gamma^{ab} p_b,\quad {dx^a\over dt}=p^a
\eeq
where $\gamma^{ab}$ is the dissipation tensor, assumed to be positive and non-degenerate. Again, if the dissipation tensor  is  the  Hessian of a convex function
\beq
\gamma_{ab}\equiv g_{ac}g_{bd}\gamma^{bd}=\nabla_a\pdr_bW,
\eeq
there are simplifications because 
\beq
{d\over dt}(\nabla^aW)+\Gamma_{bc}^a\nabla^b Wp^c=\gamma_{ab}p^b
\eeq
We define again
\beq
\tilde p_a=p_a+\pdr_aW
\eeq
to get 
\beq
{d\over dt}\pmatrix{\tilde p\cr x}=\pmatrix{\{H_1,\tilde p\}\cr \{H_1,x\}}+J\pmatrix{\{H_2,\tilde p\}\cr \{H_2,x\}}
\eeq
where
\beq
H_1=\half g^{ab}\tilde p_a\tilde p_b+V-(\nabla W)^2,\quad H_2={1\over 2\omega_2}\gamma^{ab}\tilde p_a\tilde p_b+\omega_2 W,\quad J=\pmatrix{0&-\omega_2\cr {1\over \omega_2}&0}.
\eeq
Since again $J^2=-$  it is an almost complex structure; but it may not be integrable in general. Every tangent bundle has a natural almost complex structure; $J$ is simply its translation to the co-tangent bundle $T^*Q$ using the metric $g_{ab}$ which identifies the tangent and co-tangent bundles.

Because $J$ may not be integrable, we are not able to rewrite this in terms of a complex co-ordinate $z$ in general. Nevertheless, we can think of the above equations as a generalization of hamilton's equations to a complex hamiltonian $H_1+iH_2$. 

Clearly, it is possible to quantize these system by applying the correspondence principle. Since the ideas are not very different, we will not work out the details. The  hamiltonian is:

\beq
\hat {\cal H}=\hat H+\hat H_{\rm diss},
\eeq
where
\beq
\hat H\psi=-\half\nabla^2\psi+V\psi,\quad H_{\rm diss}\psi=-\half g^{ab}\pdr_a W \pdr_b W\psi+{i\over \omega_2}\left[-{1\over 2}\nabla_a\left(\gamma^{ab}\pdr_b\psi
\right)
+\omega_2^2W
\right]+ic.
\eeq
 The imaginary constant $ic$ is chosen such that 
the ground state is stable.

\section{Acknowledgement}

I thank Andrew Jordan, Shannon Starr and Carlos Stroud for discussions
on this topic. This work is supported in part by the Department of
Energy under the  contract number 
 DE-FG02-91ER40685.

\end{document}